\documentclass[english]{llncs}
\usepackage{listings}
\usepackage{amsmath}

\title{Concurrent Constraint Machine Improvisation: Models and Implementation}
\author{Mauricio Toro}
\institute{
Universidad Eafit 
}

\begin{document}
\maketitle
\section{Introduction}


Machine improvisation 
 ``creates'' music
 either by explicit coding of rules or
by applying machine learning
methods. In this section we deal with the latter case.

An improvisation system capable of real-time must execute two
process concurrently: one to apply machine learning methods to musical sequences in order to capture prominent musical features, 
 and one to produce musical sequences stylistically consistent with the learned material.
As an example, the \textit{Concurrent Constraint Factor Oracle Model for Music Improvisation} (\textsc{ccfomi}) \cite{rad06}, based upon \textit{Non-deterministic Timed Concurrent Constraint} (\texttt{ntcc}) calculus, uses the \textit{Factor Oracle} \cite{All99}  to store the
learned sequences. 

The \textit{Factor Oracle} is a finite state
automaton that can be built efficiently online. It has two kind of
\textit{links} (i.e., transitions). \textit{Factor links} go forward and they represent at least all the factors of a
sequence (i.e., subsequences). \textit{Suffix links} go backwards and they connect
repeated patterns of the sequence.

In what follows of this section, we introduce \texttt{ntcc}, we describe \textsc{ccfomi} and its probabilistic extension, their implementations, some results and concluding remarks.

\section{The ntcc process calculus}\label{sec:ntcc}
Process calculi has been used on the modeling of interactive music systems (e.g., music improvisation and intective scores)
 \cite{is-chapter,tdcr14,ntccrt,cc-chapter,torophd,torobsc,Toro-Bermudez10,Toro15,ArandaAOPRTV09,tdcc12,toro-report09,tdc10,tdcb10,tororeport} 
 and spatially-explicit individual-based ecological systems \cite{PT13,TPSK14,PTA13,mean-field-techreport}. 

A family of process calculi is \textit{Concurrent Constraint Programming} (\texttt{ccp}) \cite{cc-chapter}, where a system is modeled in terms of variables and constraints over some variables. The constraints are contained in a common \textit{store}. There are also agents that reason about the system variables, based on  partial information (by the means of constraints).

Formally, \texttt{ccp} is based upon the idea of a \textit{constraint system (CS)}. A constraint system includes a set of (basic) constraints and a relation (i.e., entailment relation $\models$) to deduce a constraint with the information supplied by other constraints.

A \texttt{ccp} system usually includes several CSs for different variable types. 
 There are CSs for variable types such as sets, trees, graphs and natural numbers. A CS providing arithmetic relations over natural numbers is known as \textit{Finite Domain (FD)}. As an example, using a FD CS, we can deduce $pitch \neq 60$ from the constraints $pitch > 40$ and $pitch < 59$.

Although we can choose an appropriate CS to model any problem, 
in \texttt{ccp} it is not possible to delete nor change information accumulated in the store. For that reason it is difficult to perceive a notion of discrete time, useful to model reactive systems communicating with an external environment (e.g., users, lights, sensors and speakers).

\texttt{Ntcc} introduces to \texttt{ccp} the notion of discrete time as a sequence of \textit{time units}. Each time unit starts with a store  (possibly empty) supplied by the environment, and \texttt{ntcc} executes all the processes scheduled for that time unit.
In contrast to \texttt{ccp}, in \texttt{ntcc} we can model variables changing values over time. 
A variable $x$ can take different values at each time unit. To model that
in \texttt{ccp}, we have to create a new variable $x_i$ each time we change the value of $x$.


\subsection{Ntcc in multimedia interaction}

In this section we give some examples on how the computational agents of \texttt{ntcc} can be used with a FD CS.
A summary of the agents semantics can be found in Table \ref{tab:ntccagents}.

\begin{table}[!h]
  \begin{center}   
     \begin{tabular}{|ll|}
\hline
Agent & Meaning\\
\hline
\textbf{tell} $(c)$ & Adds $c$ to the current store\\
\textbf{when} $(c)$ \textbf{do} $A$  & If $c$ holds now run $A$\\
\textbf{local} $(x)$ \textbf{in} $P$  & Runs $P$ with local variable $x$\\
$A$ $\|$ $B$ & Parallel composition \\
\textbf{next} $A$ & Runs  $A$ at the next time-unit \\
\textbf{unless} $(c)$ \textbf{next} $A$  & Unless $c$ holds, next run $A$ \\
$\sum _{i \in I}$ \textbf{when} $(c_{i})$ \textbf{do} $P_{i}$  & Chooses  $P_{i}$ s.t. $(c_{i})$ holds \\ 
*$P$ & Delays P indefinitely \\ 
!$P$ & Executes P each time-unit\\
\hline
\end{tabular}    
    \caption{Semantics of \texttt{ntcc} agents.}
    \label{tab:ntccagents}
  \end{center}
\end{table}

\begin{itemize}
\item Using \textit{tell} it is possible to add constraints to the store such as $\textbf{tell} (60 < pitch_2 < 100)$, which means that $pitch_2$  is an integer between 60 and 100. 

\item \textit{When} can be used to describe how the system reacts to different events; for instance, 
$\textbf{when}$ $pitch_1=C4 \wedge pitch_2 = E4 \wedge pitch_3 = G4$ $\textbf{do}$   $\textbf{tell} (CMayor = true)$ adds the constraint 
$CMayor = true$ to the current store as soon as the pitch sequence C, E, G  has been played.

\item \textit{Parallel composition} ($\|$) makes it possible to represent concurrent processes; for instance,  
$\textbf{tell}$ $(pitch_1 = 52)$ $\|$ $\textbf{when}$ $48 < pitch_1 < 59$ $\textbf{do}$ $\textbf{tell}$ $(Instrument = 1)$ 
tells the store that $pitch_1$ is 52 and concurrently assigns the \textit{instrument} to one, since $pitch_1$ is in the desired interval.

\item \textit{Next} is useful
when we want to model variables changing over time; for instance, $ \textbf{when}$ $(pitch_1 = 60)$  $\textbf{do}$ $\textbf{next}$  $\textbf{tell}$ $(pitch_1 <> 60)$ means that if $pitch_1$ is equal to 60 in the current time unit,  it will be different from 60 in the next time unit.

\item \textit{Unless} is useful to model systems reacting when a condition is not satisfied or when the condition cannot be deduced from
the store; for instance, $\textbf{unless}$ $(pitch_1 = 60)$  $\textbf{next}$ $\textbf{tell}$ $(lastPitch <> 60)$ reacts when $pitch_1 = 60$ is false or when $pitch_1 = 60$ cannot be deduced from the store (e.g., $pitch_1$ was not played in the current time unit).

\item \textit{Star} (\textbf{*}) can be used to delay the end of a process indefinitely, but not forever; for instance, $* \textbf{tell}$ $(End = true)$. Note that to model Interactive Scores we do not use the \textit{star} agent.

\item \textit{Bang} ($\textbf{!}$)  executes a certain process every time unit after its execution; for instance, $!$$\textbf{tell}$ $(C_4 = 60)$. 

\item \textit{Sum} (\textbf{$\sum$}) is used to model non-deterministic choices; for instance, $\sum_{i \in \{48,52,55\}}$ \textbf{when} $i \in PlayedPitches$
\textbf{do} $\textbf{tell}$ $(pitch = i)$ chooses a note among those played previously that belongs to the C major chord. 

\end{itemize}

In \texttt{ntcc}, recursion can be defined (see \cite{cc-chapter}) with the form $q(x) =^{def} P_q$, where $q$ is the process name and
$P_q$ is restricted to call $q$ at most once and such call must be within the scope of a \textit{next}.
The reason of using \textit{next} is that \texttt{ntcc} does not allow 
recursion within a time unit.

The reader should not confuse a simple definition with a recursive definition; for instance, $Before_{i,j}$ $=^{def} \textbf{tell} (i \in Predecessor_j)$ is a simple
definition where the values of $i$ and $j$ are replaced statically, like a macro in a programming language.
Instead, a recursive definition such as $Clock(v)$ $=^{def} \textbf{tell} (clock=v) \| \textbf{next}\ Clock(v+1) $
is like 
a function in a programming language.

\section{Non-deterministic and probabilistic models of improvisation}
A problem of machine improvisation is to synchronize its processes. An advantage of \texttt{ntcc} is that synchronization is made declaratively by adding and deducing constraints of the store. However, \texttt{ntcc} offers little control on the recombination (i.e., combination of factors to produce new sequences). Using \textit{Probabilistic ntcc} (\texttt{pntcc}), we can use a probabilistic distribution to choose between playing a learned factor or a new sequence. In what follows we briefly describe \textsc{ccfomi} and its probabilistic extension based upon \texttt{pntcc} \cite{perez09}.


\subsection{Non-deterministic model}
\textsc{ccfomi} has three variables to represent the Factor Oracle:   $from_{k}$ is the set of labels of the 
factor links going forward from $k$, $S_{i}$ the suffix links
 from each state $i$, and  $\delta_{k,\sigma_{i}}$ the state reached from $k$
 by following a factor link labeled $\sigma_{i}$. \textsc{ccfomi} is composed by three processes: learning (\textsc{learn}), improvisation (\textsc{improv}) and synchronization (\textsc{sync}).


Process \textsc{learn} is in charge of building a representation of music: it builds up the Factor Oracle.
Process \textsc{sync} synchronizes the learning process and the user input. 
Synchronization is greatly simplified by the use of constraints. When a variable has no value, the ``when'' processes depending on it are blocked. Therefore,  \textsc{sync} is ``waiting'' until $go \geq i$: it means, the user has played note $i$, and \textsc{learn} can add a new symbol to the Factor Oracle\footnote{The condition $S_{i-1} \geq 0 $ is because the first suffix link of the Factor Oracle is equal to -1 and it cannot be followed in the simulation process (\textsc{improv}).}.  \\

$SYNC_{i}$ $\overset{def}{=}$  
\textbf{when} $S_{i-1} \geq -1 \wedge go \geq i$ \textbf{do} ($LEARN_{i}$ $\|$ \textbf{next} $SYNC_{i+1}$) \\
\hspace*{60pt}$\|$ \textbf{unless} $S_{i-1} \geq -1 \wedge go \geq i$ \textbf{next} $SYNC_{i})$ \\

The simulation process \textsc{improv} starts from state $k$ and it chooses non-deterministically
whether to output symbol $\sigma_{k}$ or to follow a suffix link $S_{k}$. When the process checks that $S_k \geq 0$, it also checks that the suffix link going backwards from $k$ exists, synchronizing \textsc{improv} with \textsc{learn}.\\

$IMPROV(k)$ $\overset{def}{=}$  \\
\hspace*{28pt} \textbf{when} $S_{k}=-1$ \textbf{do} \textbf{next} (\textbf{tell} ($out = \sigma_{k+1}$) $\|$ $IMPROV(k+1)$)  \\
\hspace*{24pt}$\|$ \textbf{when} $S_{k} \geq 0$ \textbf{do}  \textbf{next} (\\
\hspace*{40pt} (\textbf{tell} ($out = \sigma_{k+1}$)  
$\|$ $IMPROV(k+1)$)  +\\
\hspace*{42pt}$\sum _{\sigma \in \Sigma}$ \textbf{when}
$\sigma \in from_{s_{k}}$ \textbf{do} ( \textbf{tell} ($out = \sigma$)$\|$ $IMPROV(\delta_{s_{k},\sigma })$))\\
\hspace*{23pt}$\|$ \textbf{unless} $S_{k} \geq -1$  \textbf{next} $IMPROV(k)$ 

\subsection{Probabilistic model}
In the probabilistic extension of \textsc{ccfomi}, process \textsc{improv} includes a probability \textbf{$\rho$} to choose between following a factor link and a suffix link. This allows the system to control the rate of recombination.
In addition, using \texttt{pntcc} we can prove properties such as ``the system will go to a successful state with probability $q$ under $t$ discrete time-units''.
 
Authors of \texttt{pntcc} determined the value on which the model will reach an improvisation state with a
probability $q$ given 
a time bound $t$. They found out that $\rho = 0.7$ and $\rho = 0.6$ lead to a quick convergence towards an
improvisation state, whereas for lower values of $\rho$, the tendency only becomes clear after a longer time frame \cite{perez09}.

\section{Ntccrt: Implementation of ntcc and pntcc}
\textit{Ntccrt} is a framework to execute \texttt{ntcc} models \cite{toro08}.
It was shown in \cite{toro-report09} that the encoding of a 
\texttt{ntcc} process, parametrized by a Finite Domain constraint system with values between 0 and $2^{32}-1$ and a sequence of stimuli (i.e., constraints), can be translated into  sequence of \textit{Constraint Satisfaction Problems} (\textsc{csp}s). To simulate \texttt{ntcc} we do not have to solve a \textsc{csp} each time unit. Using a constraint solving library, we can use \textit{constraint propagation} to calculate the output of each time unit: Ntccrt uses  \textit{Gecode}\footnote{Gecode and constraint propagation are described in detail in \cite{tack09}.}.

\subsection{Execution of a time unit in Ntccrt}

The function $ptc(P)$ codifies a \texttt{ntcc} process $P$ (without time operators) into a constraint. It uses a \textit{reified constraint} $c \leftrightarrow b$ for each constraint guard $c$ used in a
``when'' process. In the Ntccrt, non-determinism of ``sums'' is solved with a pseudo-random choice. Reification is used because the entailment relation $\vdash$ is not implemented in most constraint solving libraries. Implementing $\vdash$ using threads that ask constantly to the constraint solving library for a value is not compatible with real time. 

\begin{center}\begin{tabular}{l l l }
$ptc(\textbf{tell}(c))$ &::= &$ c $\\

$ptc(P || Q)$ &::= &$ptc(P)\wedge ptc(Q)$\\

$ptc(\sum _{i \in I}\ \textbf{when}\ c_i \ \textbf{do}\ P_i)$ &::= &$ptc(P_k)$$, k \in \{ j, j \in I \wedge  c_j \leftrightarrow b \wedge b\}$ \\
&&\texttt{true}, otherwise \\

$ptc(\textbf{local}\ x\ \textbf{in}\ ptc(P)$ &::= &$\exists x\ ptc( P )$\\
\end{tabular}\end{center}

Given an input $c$ and a process $P$, the output of a time unit is the set of solutions of a \textsc{csp} composed by the variables that appear in $P$ with domain $[0,2^{32}-1]$ and the constraint $ptc(P)\land c$.

\subsection{Discrete time in Ntccrt}
The function $tptc(P)$ codifies a \texttt{ntcc} process into 
a pair composed by the process output in the current time unit and the process to be executed in the next time unit, the ``future of $P$''.

\begin{center}\begin{tabular}{l l l}
$tptc(\textbf{next}\ P)$ &::= &$( \texttt{true}, P )$\\
$tptc(\textbf{unless}\ c\ \textbf{next}\ P)$ &::= &(\texttt{true}, P), $c \leftrightarrow b \wedge \neg b$ \\
                        &&(\texttt{true}, \textbf{skip} ), otherwise  \\
$tptc(\textbf{!}P)$ &::= &$( c, \textbf{!}P \|Q )$  where $(c,Q) =  tptc(P) $\\
\end{tabular}\end{center} 

The function $tptc$ for ``tell'', ``when'', ``sum'', parallel composition and ``local'' is derived from the definition of $ptc$. It is described in \cite{toro-report09}.

\subsection{Programming interfaces for Ntccrt}
To execute a \textit{Ntccrt} program, we can create a stand-alone program or a plugin for either \textit{Pure Data (\textsc{pd}) or Max/MSP} \cite{maxpd}. 
Ntccrt plugins use the message passing API provided by those graphical languages to control any object in them (e.g., sound processors).

Ntccrt is written in the C++ language. Specifications can be made in Common Lisp 
 or in \textit{OpenMusic}  (a visual programming environment for music developed in Lisp) \cite{agon98}, and translated to C++. Unfortunately, there is only a parser that translates a \texttt{ntcc} specification written in Lisp or OpenMusic into C++, thus 
we cannot use Lisp functions in Ntccrt.

We want to execute \texttt{ntcc} directly in Lisp to make the power
of list processing and music languages (e.g., OpenMusic) available for Ntccrt users.
\textit{Gelisp} \cite{toro-report09a} is an Lisp-and-OpenMusic interface for the constraint solving library Gecode. In the future, we want to use Gelisp to write a Lisp version of Ntccrt.

\subsection{Probabilistic Ntccrt}
Ntccrt can also execute and verify \texttt{pntcc} models. The verification tool generates an input for a probabilistic model checker as proposed in \cite{perez09}. There is no formal encoding of \texttt{pntcc} into a sequence of \textsc{csp}s, and the implementation is still experimental.

\section{Results of machine improvisation in Ntccrt}

\textsc{Ccfomi} was executed as a stand-alone application over an Intel 2.8 \textsc{gh}z iMac using
Mac \textsc{os} 10.5.2 and Gecode 2.2.0 \cite{toro08}. Each time unit took an average of 20 ms, scheduling
around 880 processes per time unit. They simulated 300 time units and they ran each simulation 100 times in our tests.
Pachet argues in \cite{pachet02} that an improvisation system able to learn and produce sequences in less than 30 ms is appropriate for real-time interaction.
Therefore, Ntccrt is capable of real-time interaction for a simulation of \textsc{ccfomi} for at most 300 time units.

\section{Concluding remarks}
We believe that using the graphical paradigm provided by Max or \textsc{pd} is difficult and is time-demanding to synchronize processes depending on complex conditions. On the contrary, using Ntccrt, we can model such systems with a few graphical boxes in OpenMusic or a few lines in Lisp, representing complex conditions by constraints.


\let\oldbibliography\thebibliography
\renewcommand{\thebibliography}[1]{%
  \oldbibliography{#1}%
  \setlength{\itemsep}{0pt}%
}

\bibliographystyle{abbrv}
{\scriptsize
\bibliography{biblio}
}

\end{document}